\documentclass[apj]{emulateapj}
\slugcomment{{\sc Accepted to ApJ:} July 1, 2016}
\usepackage{url}
\usepackage[utf8]{inputenc}
\usepackage{fancyref}
\usepackage{hyperref}
\hypersetup{colorlinks,breaklinks,
            urlcolor=[rgb]{0.2,0.2,0.2},  
            linkcolor=[rgb]{0.843,0.098,0.110},  
            citecolor=[rgb]{0,0.4,1.0}}  
\usepackage{amssymb,amsmath}
\usepackage{natbib}
\usepackage{epstopdf}
\usepackage{graphicx}
\usepackage{floatrow}
\usepackage{rotating}
\usepackage{color}

\newcommand{\teff}{${T}_{\mathrm{eff}}$}
\newcommand{\logg}{$\log{g}$}
\newcommand{\msun}{$M_{\odot}$}

\newcommand{\muhz}{$\mathrm{\mu}$Hz}

\newcommand{\Kep}{\emph{Kepler}}
\newcommand{\Ktwo}{\emph{K2}}

\shorttitle{Outbursts in the Coolest Pulsating White Dwarfs}
\shortauthors{Bell et al.}

\begin{document}
\title{Outbursts in Two New Cool Pulsating DA White Dwarfs}
\author{Keaton J. Bell\altaffilmark{1,2}, J. J. Hermes\altaffilmark{3,4}, M. H. Montgomery\altaffilmark{1,2}, N.~P.~Gentile~Fusillo\altaffilmark{5}, R.~Raddi\altaffilmark{5}, B.~T.~G\"{a}nsicke\altaffilmark{5}, D.~E.~Winget\altaffilmark{1,2}, E.~Dennihy\altaffilmark{3}, A.~Gianninas\altaffilmark{6}, P.-E.~Tremblay\altaffilmark{5}, P.~Chote\altaffilmark{5}, and K. I. Winget\altaffilmark{1,2}}
\altaffiltext{1}{Department of Astronomy, University of Texas at Austin, Austin, TX\,-\,78712, USA}
\altaffiltext{2}{McDonald Observatory, Fort Davis, TX\,-\,79734, USA}
\altaffiltext{3}{Department of Physics and Astronomy, University of North Carolina, Chapel Hill, NC\,-\, 27599, USA}
\altaffiltext{4}{Hubble Fellow}
\altaffiltext{5}{Department of Physics, University of Warwick, Coventry\,-\,CV4~7AL, UK}
\altaffiltext{6}{Homer L. Dodge Department of Physics and Astronomy, University of Oklahoma, 440~W.~Brooks~St., Norman, OK\,-\,73019, USA}

\email{keatonb@astro.as.utexas.edu}

\begin{abstract}
The unprecedented extent of coverage provided by \Kep\ observations recently revealed outbursts in two hydrogen-atmosphere pulsating white dwarfs (DAVs) that cause hours-long increases in the overall mean flux of up to 14\%.
We have identified two new outbursting pulsating white dwarfs in \Ktwo, bringing the total number of known outbursting white dwarfs to four. EPIC\,211629697, with \teff\ = $10{,}780 \pm 140$\,K and \logg{} = $7.94 \pm 0.08$, shows outbursts recurring on average every 5.0\,d, increasing the overall flux by up to 15\%. EPIC\,229227292, with \teff\ = $11{,}190 \pm 170$\,K and \logg\ = $8.02 \pm 0.05$, has outbursts that recur roughly every 2.4\,d with amplitudes up to 9\%. We establish that only the coolest pulsating white dwarfs within a small temperature range near the cool, red edge of the DAV instability strip exhibit these outbursts.

\end{abstract}


\section{Introduction}

White dwarf stars are the remnant products of 97\% of Galactic stellar evolution. About 80\% of white dwarfs spectroscopically display atmospheres dominated by hydrogen \citep[DA;][]{Tremblay2008}.
Convective driving \citep{Brickhill1991,Goldreich1999} of nonradial gravity-mode pulsations \citep{Robinson1982} in DA white dwarfs between 12,500~K $>$ \teff\  $>$ 10,600~K \citep[for typical \logg\ $\approx 8.0$;][]{Tremblay2015} causes these objects to appear photometrically variable. The frequencies of photometric variability are eigenfrequencies of these stars as physical systems, providing a powerful tool for studying their interior structures \citep[see reviews by][]{Winget2008,Fontaine2008,Althaus2010}.

The \Kep\ spacecraft has provided unrivaled monitoring of pulsating white dwarfs, both in its original mission and during the two-reaction-wheel mission, \Ktwo\ \citep{Howell2014}. The first and longest-observed pulsating DA white dwarf (DAV) known to lie within the original mission field is KIC\,4552982 \citep[WD~J191643.83+393849.7;][]{Hermes2011}. This target was observed nearly continuously every minute for more than 1.5\,yr.  Unexpectedly, these data revealed at least 178 brightness increases that recurred stochastically on an average timescale of 2.7\,d. The events increased the total flux output of the star by $2-17$\% and lasted $4-25$\,hr \citep{Bell2015}.

\citet{Hermes2015} described a second DAV to display similar outburst behavior: PG\,1149+057, observed in \Ktwo\ Campaign 1. These outbursts caused the mean flux level to increase by up to 14\%, which would correspond to a nearly 750\,K global increase in the stellar effective temperature, with a recurrence timescale of roughly 8\,d and a median duration of 15\,hr. Mean pulsation frequencies and amplitudes were both observed to increase in this star during outbursts, and the combined flux enhancement from outbursts and high amplitude pulsations reached as high as 45\%. 
The outbursts affect the pulsation properties of PG\,1149+057, and \citet{Hermes2015} unambiguously ruled out a close companion or a line-of-sight contaminant as the source of this phenomenon.

Spectroscopic effective temperatures place both of these white dwarfs very near to the empirical cool edge of the DAV instability strip---the boundary below which pulsations have not been detected in white dwarfs. While nonadiabatic pulsation codes successfully reproduce the observed hot edge of the DAV instability strip, they typically predict a cool edge thousands of Kelvin below what is observed \citep[e.g.,][]{VanGrootel2012}. The discovery of a new astrophysical phenomenon that operates precisely where our models are discrepant with observations suggests that the continued discovery and study of cool outbursting DAVs may inform fundamental improvements to the theory of stellar pulsations.

In this paper we present the identification of two new outbursting DAVs that were observed by \Ktwo\, along with one candidate outburster. EPIC\,211629697 was observed at short cadence in \Ktwo\ Campaign 5 and EPIC\,229227292 in Campaign 6. Both stars are qualitatively similar in outburst and pulsational properties to the two previously published objects. We characterize these stars in Sections 2 and 3, respectively. Additionally, we inspect the light curves of the hundreds of other white dwarfs already observed by \Ktwo\ in Section 4, and describe a candidate single outburst in the long-cadence data of EPIC\,211891315. We summarize the current members of the outbursting class of DAV and discuss possible physical mechanisms and outburst selection effects in Section 5.

\begin{figure*}
  \centering
  \includegraphics[width=0.99\columnwidth]{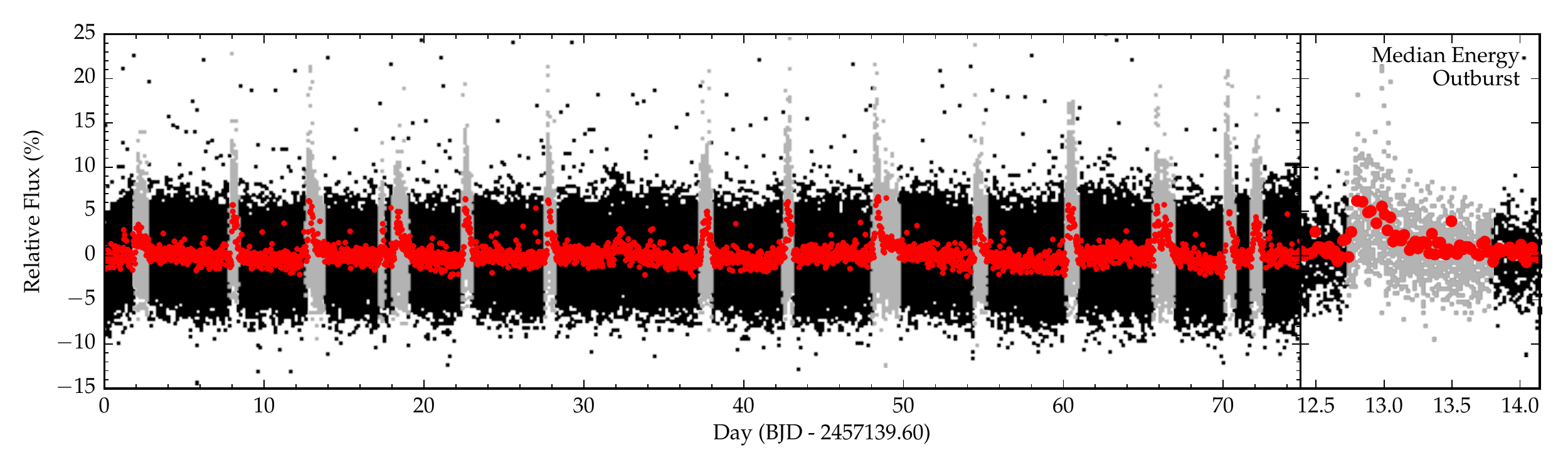}
  \caption{\emph{Left:} The \Ktwo{} Campaign 5 light curve of EPIC\,211629697. The short-cadence data are displayed in black (during quiescence) and gray (during the 15 detected outbursts). The long-cadence data are shown in red. \emph{Right:} A detailed view of the outburst of median energy (see text). The units on the x-axes are the same in both panels.  The scales of the y-axes are identical, with greater apparent scatter in the left panel due only to the overlap of points.}
  \label{fig:oDAV3lc}
\end{figure*}

\begin{figure}
  \centering
  \includegraphics[width=0.99\columnwidth]{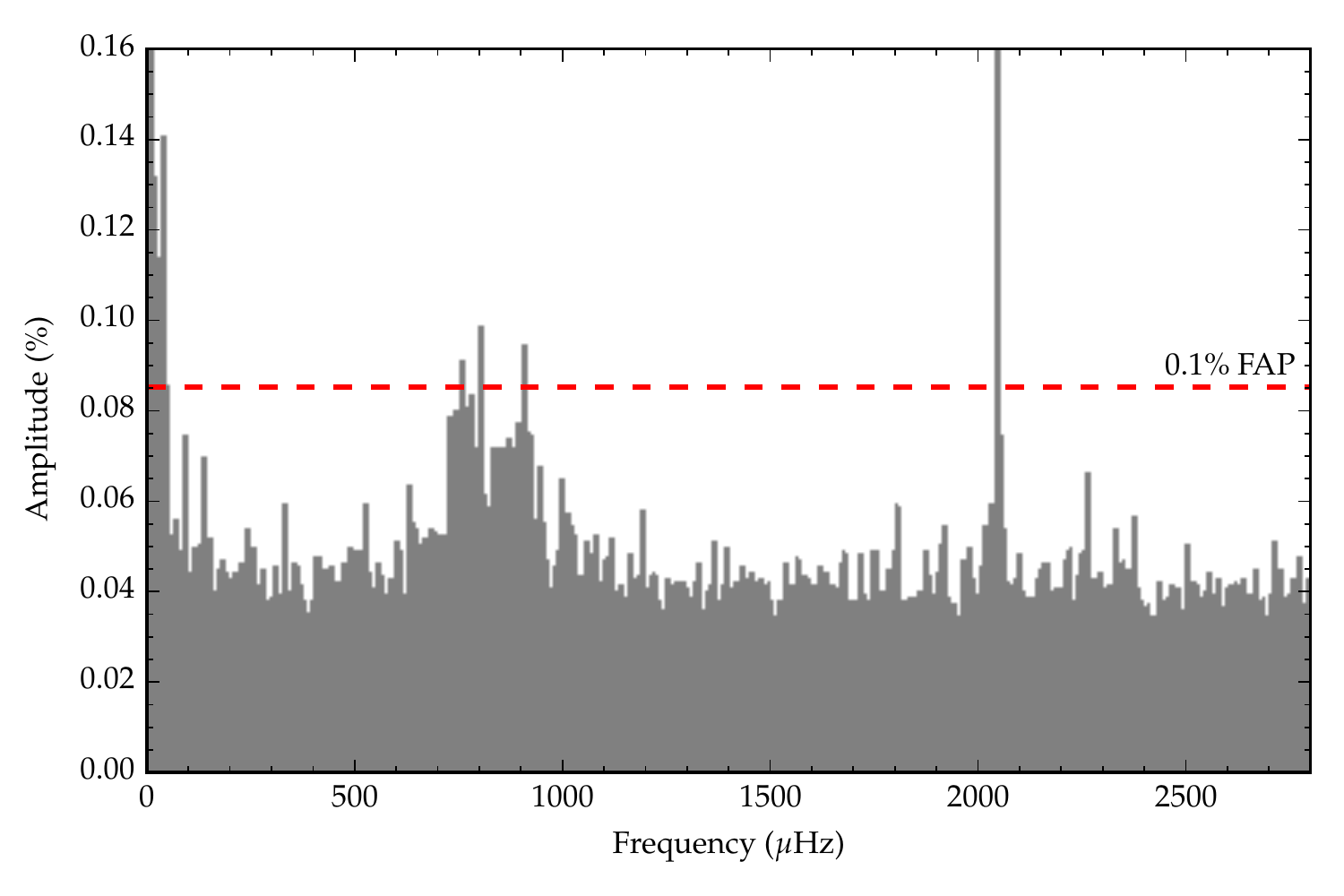}
  \caption{Fourier transform of the entire \Ktwo\ light curve of EPIC\,211629697, including outbursts. The dashed line gives the 0.1\% False Alarm Probability (FAP) significance threshold for a single peak determined from bootstrapping (see text). The peak at 2053.514\,\muhz\ (486.97\,s) and 3 frequencies in the range 764--913\,\muhz\  (1095--1309\,s) reach amplitudes that exceed this significance threshold. We discard all low-frequency peaks below 100\,\muhz\ (see text).}
  \label{fig:oDAV3ft}
\end{figure}

\section{The Third Outburster: EPIC 211629697}
\label{sec:oDAV3}

We targeted the DA white dwarf EPIC\,211629697 ($K_p=18.4$ \,mag, SDSSJ\,084054.14+145709.0) for short-cadence (58.8\,s) monitoring as part of our \Ktwo\ Guest Observer program searching for candidate pulsating white dwarfs (GO5043). The effective temperature from an automated fit to a spectrum from the Sloan Digital Sky Survey (SDSS; \citealt{Kleinman2013}) put this white dwarf within the empirical DAV instability strip, although it was not previously known to pulsate. We have updated the one-dimensional atmospheric parameters from the SDSS spectrum by refitting these data with the latest atmosphere models described in \citet{Tremblay2009}, which use the ML2$/\alpha = 0.8$ prescription of the mixing-length theory, and corrected the values to compensate for the three-dimensional dependence of convection \citep{Tremblay2013}. We find this white dwarf has \teff\ = $10{,}780 \pm 140$\,K and \logg{} = $7.94 \pm 0.08$, corresponding to a mass of $0.57\pm0.04$ \msun .

The light curve was obtained at short cadence in \Ktwo\ Campaign 5, spanning 2015~April~27 02:25:19 UT to 2015~July~10 22:36:12 UT. The raw pixel-level data were extracted and detrended using the pipeline described in \citet{Armstrong2015}, which corrects for attitude readjustments of the spacecraft on multiples of every 5.9\,hr. Our extraction uses a fixed pattern of 4 pixels centered on the target. Despite the large \Kep\ pixels, there is no contamination from nearby stars in our extraction.

Subsequently, we clip the light curve of 77 outliers that lie $>$$4\sigma$ below or $>$$6\sigma$ above the local median flux (calculated for 30\,m bins along the light curve, where $\sigma$ is the standard deviation of flux measurements), leaving $107{,}682$ observations over 74.84\,d. We then subtract out a 6th-order polynomial fit to the full light curve to mitigate some of the long-term instrumental systematics. 

In addition to this short-cadence light curve, we also analyze the pre-search data conditioned long-cadence light curve produced by the \Kep\ Guest Observer office \citep{Twicken2010}.

The reduced short- and long-cadence \Ktwo\ light curves of EPIC\,211629697 are presented in Figure~\ref{fig:oDAV3lc}. We display the Fourier transform (FT) of the entire short-cadence light curve (including outbursts) in Figure~\ref{fig:oDAV3ft}.

\subsection{Outbursts}

We detect a total of 15 outbursts that cause significant brightness enhancements in the \Ktwo\ observations of EPIC 211629697.  These outbursts are identified by an automatic algorithm wherever two consecutive points in the long-cadence light curve exceed 3 times the overall standard deviation measured in the light curve.  We define the start and end times of each outburst as where the long-cadence light curve first crosses the median measured flux level immediately before and after these significantly high flux excursions.  We mask out these regions of the light curve and recompute the overall standard deviation, repeating the candidate outburst search until no new features are flagged. For EPIC\,211629697, this process yields 16 candidate outburst detections.  We scrutinize these candidate events in both the long- and short-cadence light curves, determining one candidate to be a spurious detection that is not present in the short-cadence data. The remaining 15 outbursts are highlighted in the left panel of Figure~\ref{fig:oDAV3lc}. 

The outbursts increase the mean stellar brightness by between $6-15$\% (defined as the greatest median value of any 6 consecutive points in the short-cadence light curve during each outburst), and the mean time between consecutive outbursts is roughly 5.0\,d. The median measured outburst duration is 16.3\,hr.

We characterize the excess energy of the outbursts in the \Kep\ bandpass by calculating their equivalent durations (integrated excess flux in the short-cadence light curve, similar to a spectroscopic equivalent width), as described in \citet{Bell2015}. Equivalent durations equal the amount of time that the white dwarf would have to shine in quiescence to output as much flux in the \Kep\ bandpass as the flux excess measured during these outbursts.  The median equivalent duration that we measure for an outburst in EPIC\,211629697 is 21\,min (this outburst, the third, is displayed in better detail in the right panel of Figure~\ref{fig:oDAV3lc}). The maximum measured equivalent duration is 35\,min.

These equivalent durations can be converted to approximate outburst energies by making a few simplifying assumptions: that the flux enhancement from an outburst is the same at all wavelengths, and that outbursts are isotropic.  We calculate the bolometric luminosity of EPIC 211629697 using the Stefan–Boltzmann law and the parameters of the model that yielded the best fit to the SDSS spectrum \citep{Tremblay2009,Fontaine2001}. This value of $L_{\rm bol} = 8.36\times 10^{30}$\,erg\,s$^{-1}$ is the scaling factor between equivalent duration and outburst energy, yielding a median outburst energy of  $1.1\times 10^{34}$\,erg, and a maximum energy of $1.8\times 10^{34}$\,erg.

We note that our ability to detect outbursts is limited by the signal-to-noise of the light curve. EPIC\,211629697 is relatively faint, at $K_p=18.4$\,mag, and the final threshold for two consecutive points in the long cadence light curve to flag an outburst in our detection scheme is set to 2.39\%.  It is possible that this star undergoes smaller-amplitude outbursts that we are unable to detect in this data set.  The summary characterization of outbursts given above represents the detected outbursts and may not be directly comparable to outbursts from other DAVs that were observed with different photometric precision.

\subsection{Pulsations}
\label{sec:oDAV3pulsations}

\begin{figure*}
  \centering
  \includegraphics[width=0.99\columnwidth]{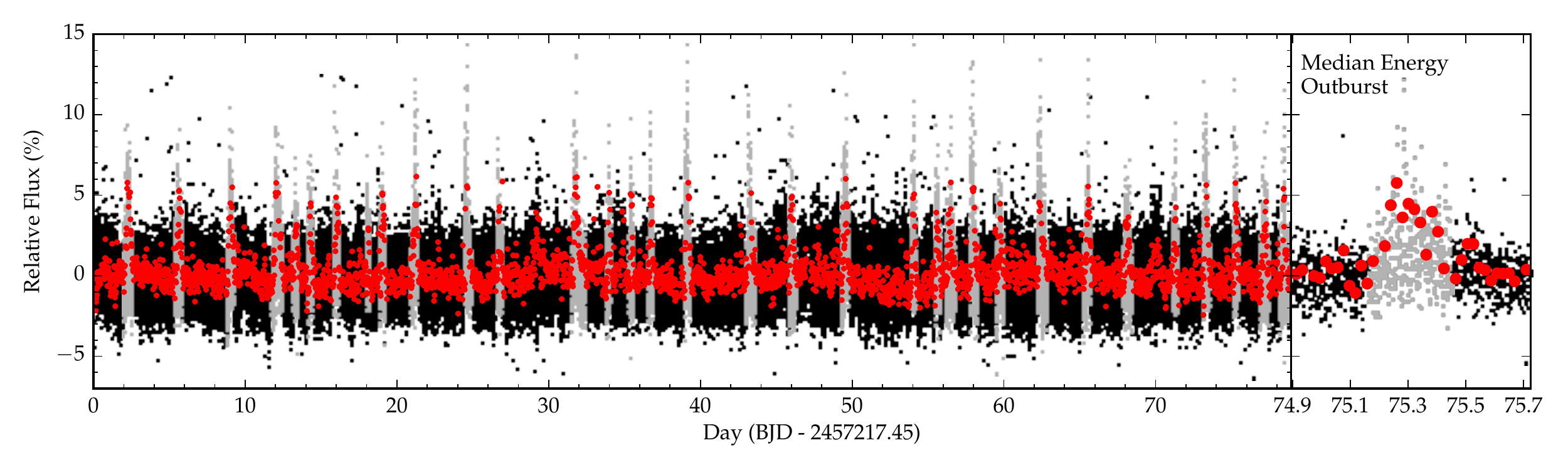}
  \caption{The \Ktwo\ Campaign 6 light curve of EPIC\,229227292. The short cadence data are presented in black (in quiescence) and gray (in outburst), with the long cadence data in red. \emph{Left:} The full light curve featuring 33 significant outbursts. \emph{Right:} A detailed view of the outburst of median energy, with the same y-axis scale and x-axis units.}
  
  \label{fig:oDAV4lc}
\end{figure*}

\begin{figure}
  \centering
  \includegraphics[width=0.99\columnwidth]{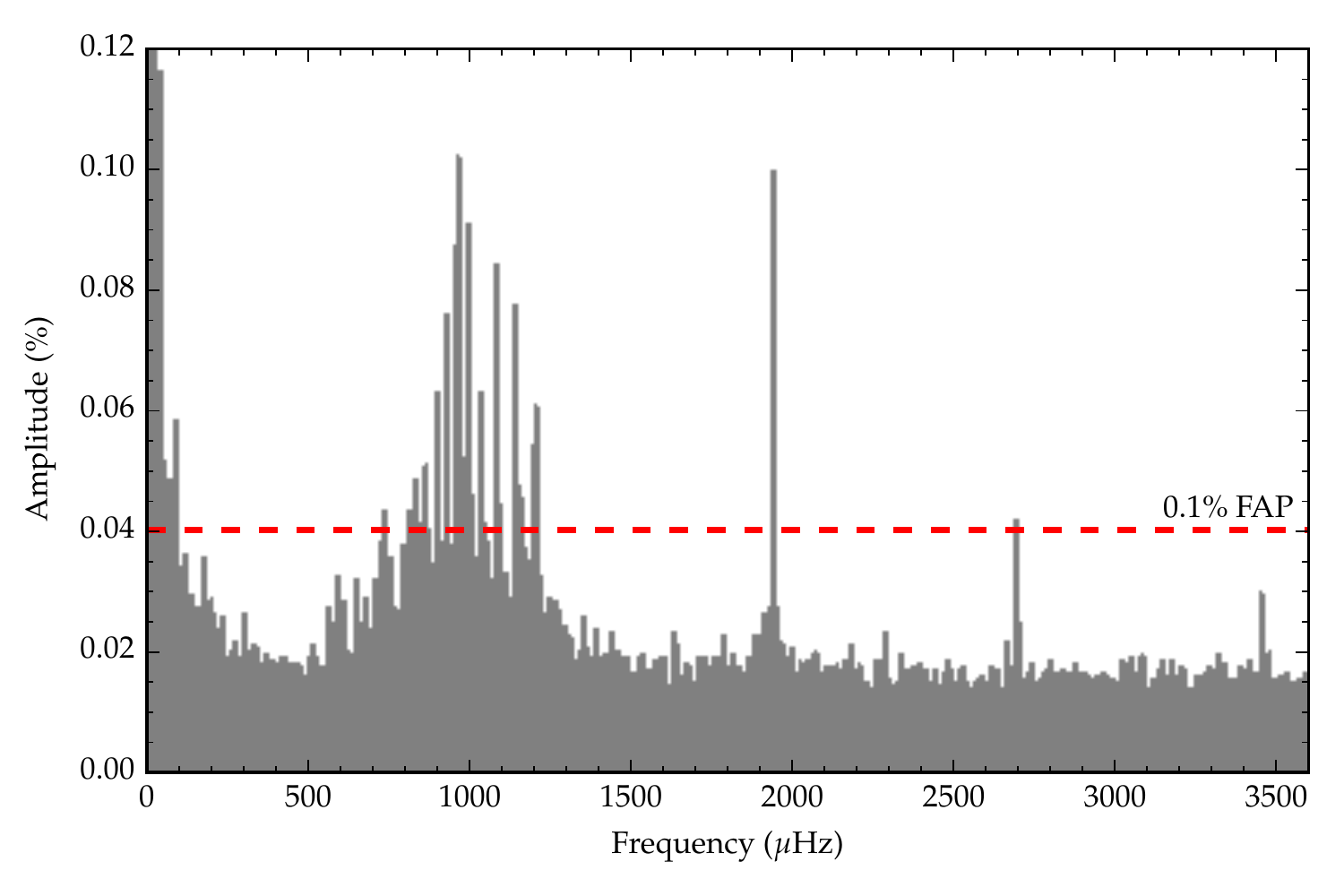}
  \caption{FT of the \Ktwo\ light curve of EPIC\,229227292. We detect several peaks in the range $800-1250$\,\muhz\ ($800-1250$\,s), and two peaks at 1945.048 and 2697.423\,\muhz\ ($\approx$ 371 and 514\,s) that exceed our 0.1\% FAP significance threshold (dashed line).  The peak just below the significance threshold at 3456.41\,\muhz\ (289\,s) is highly suggestive of being astrophysical signal (see text).}
  \label{fig:oDAV4ft}
\end{figure}

We detect significant but low-amplitude pulsations in EPIC\,211629697, and show the FT in Figure~\ref{fig:oDAV3ft}. Asteroseismic analysis is beyond the scope of this paper and will be addressed in future publications, but we do characterize the pulsation frequencies generally in this observationally-focused work.  All FTs in this paper are oversampled by a factor of 20.

We use a bootstrap method to identify statistically significant signals in the FT. After prewhitening the light curve of known instrumental artifacts that are harmonics of the long-cadence sampling rate \citep{Gilliland2010}, we shuffle the points in the light curve and recalculate the FT $10{,}000$ times \citep{Bell2015}. This shuffling preserves the exact time sampling of the original light curve, but destroys the coherence of any underlying signals. We treat the FTs of the shuffled light curves as proxies for the underlying noise spectrum, though this yields a conservative estimate for the typical noise level because the photometric scatter is inflated by the mixed-in signal. For this reason, we understate our true confidence in signal that exceeds our significance criterion.

When we consider the full set of $10{,}000$ noise simulations, we find that the peak value anywhere in the FT---out to the Nyquist frequency---exceeds a value of 0.0853\% in fewer than 1/1000 runs. We indicate this value with a dashed line in Figure~\ref{fig:oDAV3ft} as the 0.1\% false alarm probability (FAP) threshold for any individual peak in the FT.  

Besides the noise at low-frequency (below 100\,\muhz) that results from both the presence of outbursts in the light curve and residual systematics of the \Ktwo\ photometry, including the $\sim5.9$\,hr thruster firing timescale, there are numerous signals resulting from stellar pulsations that exceed this significance threshold in the FT.  The highest peak is the sharp signal at $2053.514 \pm 0.007$\,\muhz\ that reaches an amplitude of $0.161\pm 0.014$\% \citep[formal analytical uncertainties calculated following \citealt{Montgomery1999} with the {\sc Period04} software;][]{Lenz2004}.  The FT also reveals 3 significant resolved frequencies in the range 764--913\,\muhz .  This cluster of significant frequencies likely corresponds to a sequence of pulsational power bands---modes that are not strictly coherent over the course of observations---as were observed in the previous two cases of outbursting cool DAVs.  The signal-to-noise of this data set is not sufficient for easily identifying an exhaustive list of individual frequencies associated with pulsational eigenmodes of this star, so we characterize generally the pulsational properties of this star as consisting of bands of power in the range 764--913\,\muhz\ with a more stable mode at the higher frequency, 2053.514\,\muhz. The actual frequency range of excited pulsational modes in the power band region is likely broader than the formally significant range reported, which is limited by the photometric signal-to-noise and baseline of observations.

\begin{figure*}
  \centering
  \includegraphics[width=0.99\columnwidth]{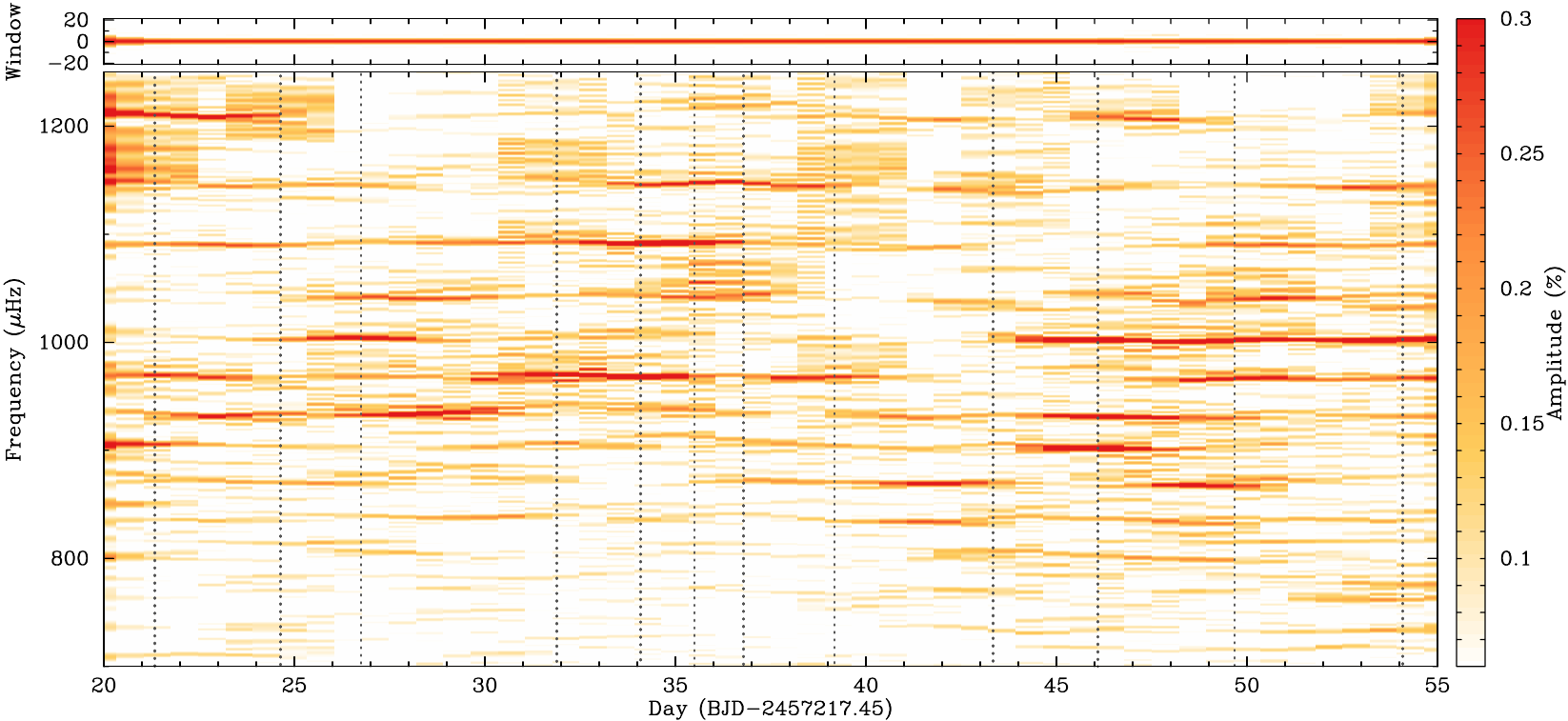}
  \caption{Running FT of EPIC\,229227292, showing how the amplitude of pulsations changes in relation to outbursts. Vertical dotted lines mark the times of maximum brightness during the detected outbursts in Figure~\ref{fig:oDAV4lc}; we use a 3-day sliding window, which smears the events. We note that the rapid growth/decay of power in individual modes commonly coincides with a detected outburst (e.g., the dropping out of power near 1209\,\muhz\ that immediately follows the outburst near Day 24). This strongly suggests that the observed pulsations respond to outbursts in EPIC\,229227292, as was observed in PG\,1149+057 \citep{Hermes2015}.}
  \label{fig:oDAV4runningFT}
\end{figure*}

\section{The Fourth Outburster: EPIC 229227292}
\label{sec:oDAV4}

We targeted EPIC\,229227292\footnote{This target received a duplicate EPIC identifier, and is also cataloged as EPIC\,229228124.} ($K_p=16.7$\,mag, ATLASJ\,134211.62$-$073540.1) for short-cadence \Ktwo\ monitoring using an early data release of the VST/ATLAS survey, which is a deep $ugriz$ photometric survey of the southern hemisphere \citep{Shanks2015}. Based on its high reduced proper motion and $ugr$ colors, we considered the object a high-probability white dwarf near the DAV instability strip and proposed observation in \Ktwo\ Campaign 6 (proposal GO6083).

As with EPIC\,211629697, we extracted and detrended the short-cadence light curve using the pipeline described in \citet{Armstrong2015} and use the long-cadence light curve from the \Kep\ Guest Observer office. We clipped the short-cadence data of outliers $>$$4\sigma$ below or $>$$8\sigma$ above the local median (32 total, with the higher threshold above the median value to preserve astrophysical signal), leaving $113{,}635$ individual observations over 78.93\,d. Our final light curves, spanning 2015~July~13 22:54:00 UT to 2015~September~30 21:08:31, are displayed in Figure~\ref{fig:oDAV4lc}. The FT of the entire short-cadence light curve, including the data in outburst, is shown in Figure~\ref{fig:oDAV4ft}.

One complication in our extraction came from the presence of charge bleed in the \Ktwo\ target pixels caused by the saturation of naked-eye M dwarf, 82\,Virginis (a.k.a. the known variable $m$\,Vir, J134136.78$-$084210.7), which falls roughly 1\,deg south of EPIC\,229227292. Our $3\times3$ pixel extraction aperture centered on the white dwarf excludes this hot column. We have also ensured that this object does not contaminate our photometry by inspecting the light curve extracted from only the top and bottom two pixels of this charge bleed column, where we do not see evidence of brightening events from $m$\,Vir on the same timescale as the outbursts.  As we discuss in Section \ref{sec:oDAV4pulsations}, the outbursts affect the pulsations, confirming that these brightening events are occurring on the white dwarf.

\subsection{Outbursts}

We identify 33 significant outbursts in the long-cadence light curve of EPIC 229227292 with the same automated method as used for EPIC 211629697 (after discarding four spurious detections that are not corroborated by the short-cadence data).  These outbursts are highlighted in the left panel of Figure~\ref{fig:oDAV4lc}. The outbursts reach amplitudes of 4--9\% (the peak local median of 6 consecutive points in the short cadence light curve), with a median 
duration of 10.2\,hr before returning to quiescence. The mean time between outbursts is 2.4\,d.

The detected outbursts have equivalent durations (proportional to outburst energy in the \Kep\ bandpass) between $2.6-12$\,min, with a median of 5.8\,min. Following the same approach as for EPIC 211629697 and using the spectroscopic and model parameters determined in Section~\ref{sec:oDAV4spec}, we convert these to total outburst energies in the range $1.4-6.3\times 10^{33}$\,erg, with a median energy of $3.1\times 10^{33}$\,erg.  The outburst of median equivalent duration is displayed in the right panel of Figure~\ref{fig:oDAV4lc}.

\subsection{Pulsations}
\label{sec:oDAV4pulsations}

The FT of the full EPIC 229227292 light curve (including outbursts) in the region of astrophysical power is presented in Figure~\ref{fig:oDAV4ft}. We use the same bootstrap approach as before to calculate a 0.1\% FAP significance threshold of 0.0403\% for single peaks in the FT.  Again, we do not believe any power $<100$\,\muhz\ arises directly from stellar pulsations.

Owing to higher photometric signal-to-noise for this brighter object, our significance criterion is much lower and we can discern more details of the pulsational signatures in the FT.  We detect at least 11 wide bands of pulsational power clustered in the range $800-1250$\,\muhz, with two relatively stable pulsation modes at higher frequencies: $1945.048 \pm 0.005$ and $2697.423 \pm 0.013$\,\muhz\ \citep[analytical uncertainties;][]{Montgomery1999}.  The peak at $3456.41\pm 0.02\,\mu$Hz is also highly suggestive, rising to a signal-to-noise of 4.95 (defined as the ratio of the peak amplitude to the local mean amplitude in the FT, $\langle {\rm A}\rangle$), but does not meet our adopted significance criterion.  With our significance threshold being a conservative estimate, it is difficult to assess the precise likelihood of this frequency belonging to a pulsation mode in the star, but we mention it as a tentative astrophysical signal.  The approach to determining detection thresholds in the FTs of \Ktwo\ short-cadence observations of \citet{Baran2015} assigns a confidence of $\approx 90$\% to peaks with this signal-to-noise ratio, so this is likely the highest frequency pulsation mode observed in an outbursting DAV so far.

The high signal-to-noise of the EPIC 229227292 data also enables us to explore changes in the pulsations on shorter timescales through the running FT, displayed for the 20th to 55th day of observations in Figure \ref{fig:oDAV4runningFT}.  This shows the evolution of the FT as calculated for a three-day sliding window on the light curve in the region of pulsational power bands.  Individual mode amplitudes are observed to grow and decay dramatically on the timescale of days.  The times of detected outburst peaks are indicated with vertical dotted lines.  We note that the outbursts coincide in many cases with the sudden growth or decay of mode amplitudes, suggesting that the outbursts play a role in redirecting pulsational energy, and that they at least have some effect on the pulsations.

\subsection{Spectroscopy}
\label{sec:oDAV4spec}

No spectroscopy of EPIC\,229227292 existed previous to this work. After discovering pulsations, we followed up this white dwarf using the Goodman spectrograph on the 4.1\,m SOAR telescope \citep{Clemens2004}. We obtained 6$\times$180\,s exposures taken consecutively on 2016~February~15, covering roughly $3700-5200$\,\AA\ with a dispersion of 0.84\,\AA\ pixel$^{-1}$. Using a 3\arcsec\ slit, our resolution was seeing limited; seeing was steady around 1.1\arcsec\ during our observations, yielding a roughly 3.2\,\AA\ resolution. Each exposure had a signal-to-noise (S/N) of roughly 17 per resolution element in the continuum at 4600\,\AA, for an overall S/N $\simeq41$. Including overheads, our observations span roughly 18.5\,min, covering at least one pulsation cycle for most oscillations. 

We processed the images using the {\sc starlink} packages {\sc figaro} and {\sc kappa}, and optimally extracted the spectra \citep{Horne1986} using the {\sc pamela} package \citep{Marsh1989}. Wavelength and flux calibration were performed with an FeAr lamp and the spectrophotometric standard GD~71, using the {\sc molly} package\footnote{\href{http://www.warwick.ac.uk/go/trmarsh}{http://www.warwick.ac.uk/go/trmarsh}}.

\begin{figure}[b]
  \centering
  \includegraphics[width=0.7\columnwidth]{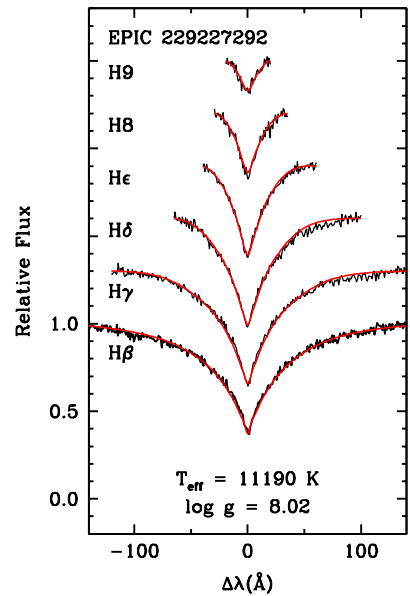} 
  \caption{The best-fit atmosphere model (red) plotted over the average of the SOAR spectra (black) of EPIC 229227292 shows the agreement in the Balmer lines.  Each spectral line is offset vertically by a factor of 0.3 for clarity.}
  \label{fig:spec}
\end{figure}

\begin{figure*}[t]
  \centering
  \includegraphics[width=0.980\columnwidth]{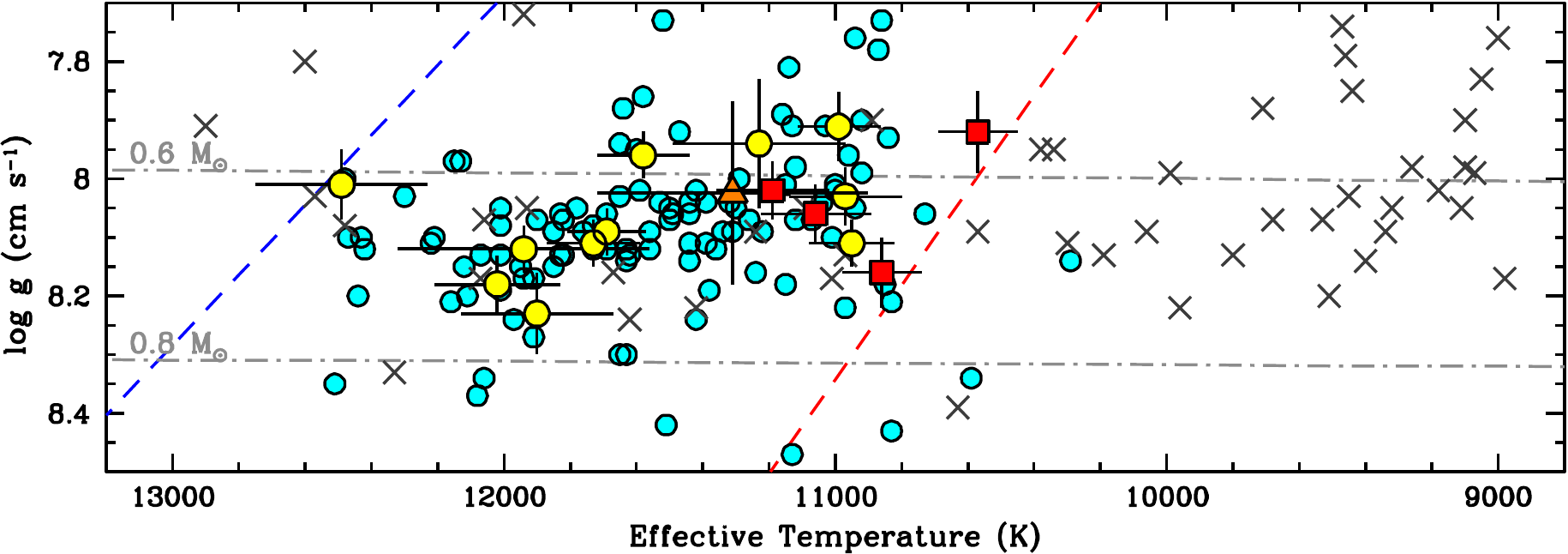}
  \caption{The location of the four known outbursting DAVs (red squares) in \logg --\teff{} parameter space, as well as the candidate EPIC\,211891315 (orange triangle; see Section~\ref{sec:oDAV5}). The crosses show white dwarfs observed by \Ktwo\ which do not show outbursts (Table~\ref{tab:novoutb}), confining this outburst phenomenon to the coolest pulsating white dwarfs, between roughly $10{,}600$\,K and $11{,}200$\,K. Previously known DAVs are shown in cyan circles \citep{Tremblay2011,Gianninas2011}, and non-outbursting pulsating white dwarfs observed with \Kep\ are shown in yellow (with error bars; see Table~\ref{tab:novoutb}). The empirical instability strip is demarcated with blue and red dashed lines \citep{Tremblay2015}. All atmospheric parameters have been corrected for the 3D-dependence of convection \citep{Tremblay2013}.  The dash-dotted gray lines mark evolutionary cooling tracks for 0.6 \msun\ and 0.8 \msun\ white dwarfs \citep{Fontaine2001}.}
  \label{fig:paramspace}
\end{figure*}

We fit each individual spectrum with a set of one-dimensional pure-hydrogen-atmosphere models and fitting procedure described in \citet{Gianninas2011} and references therein, which use ML2/$\alpha = 0.8$. We found the weighted mean of these individual exposures, and used the 3D convective corrections of \citet{Tremblay2013} to determine the atmospheric parameters of EPIC\,229227292 to be \teff\ = $11{,}190 \pm 170$\,K and \logg\ = $8.02 \pm 0.05$, corresponding to a mass of $0.62\pm0.03$ \msun\ \citep{Fontaine2001}.  The best-fit atmosphere model is plotted over the spectroscopic data in Figure~\ref{fig:spec} to demonstrate the quality of the fit.

These atmospheric parameters indicate that EPIC\,229227292 is consistent with being the hottest outbursting DAV discovered so far, but this white dwarf is still located close to the cool edge of the DAV instability strip. We estimate that this white dwarf was in outburst for 19\% of the 78.93\,d that it was monitored by \Ktwo. There is thus a roughly one-in-five chance that some of our spectroscopy was taken in outburst, which may contribute to the relatively high \teff\ measured.

\section{A Wider Search for Outbursts}
\label{sec:search}

As part of a search for transits and rotational variability of stellar remnants, \Ktwo\ has already observed several hundred white dwarfs in the first six campaigns, mostly at long cadence with exposures taken every 29.4\,min. These targets have been proposed by a number of different teams\footnote{The white dwarfs described in this section were proposed for \Ktwo\ observations by teams led by M.~Kilic, M.~R.~Burleigh, Seth~Redfield, Avi~Shporer, Steven~D.~Kawaler, and our team.}, leading to the discovery of the first transits of a white dwarf: the object WD\,1145+017\footnote{WD\,1145+017 was proposed jointly by teams led by M.~Burleigh, M.~Kilic, and Seth~Redfield, searching for transits.}, observed in \Ktwo\ Campaign\,1, is transited every $\sim$4.5\,hr by a disintegrating minor planet \citep{Vanderburg2015}.

In all, more than 300 spectroscopically confirmed DA white dwarfs have been observed by \Ktwo\ through Campaign\,6, spanning temperatures from 4800\,K up to $100{,}000$\,K. These light curves provide a unique opportunity to immediately constrain the temperature distribution of outbursting white dwarfs. Our automatic detection algorithm considers only the long-cadence light curves, demonstrating that these observations are sufficient to detect outbursts since the events typically have durations of many hours (see Figures \ref{fig:oDAV3lc}, \ref{fig:oDAV4lc}).

We put these outbursting and non-outbursting white dwarfs into context in Figure~\ref{fig:paramspace}. We focus in detail on a subset of 52 white dwarfs that have effective temperatures within 2000\,K of $10{,}900$\,K, roughly the mean effective temperature of the first four outbursting DAVs. In all cases, the atmospheric parameters have been obtained from the SDSS spectra using the models described in \citet{Tremblay2011} and ML2$/\alpha = 0.8$, with the exception of EPIC\,203705962 \citep{Kawka2006} and EPIC\,212564858 \citep{Koester2009}. All parameters have been corrected for the 3D-dependence of convection \citep{Tremblay2013} and are listed in Table~\ref{tab:novoutb}.

The long-cadence light curves for this subsample were obtained from the Mikulski Archive for Space Telescopes (MAST). In each case, we have either used extracted and de-trended light curves from the pipeline described in \citet{Vanderburg2014} (VJ) or, from Campaign 3 and onward, the pre-search data conditioned light curves produced by the \Kep\ Guest Observer office (GO). Both pipelines mitigate for attitude corrections from \Ktwo\ thruster firings, but in slightly different ways, and we include in Table~\ref{tab:novoutb} which pipeline we use for our outburst analysis. We have ensured that the apertures used enclose only the white dwarf target.

Many of these targets are very faint, with $K_p>19.0$\,mag, which is why we did not propose short-cadence observations of those with temperatures inside the empirical instability strip. However, \Ktwo\ has proven itself stable enough to deliver useful long-cadence photometry on these faint targets; a $K_p=19.2$\,mag target typically has roughly 1.2\% r.m.s. scatter, which increases to roughly 3\% for a $K_p=19.5$\,mag target.
We assign limits on a non-detection of outbursts in these targets in Table~\ref{tab:novoutb}. These limits were calculated by comparing the highest three consecutive points in the long-cadence light curves with the overall standard deviation of flux measurements ($\sigma$). The long-cadence light curves of the known outbursting DAVs all show multiple occurrences of at least three consecutive points exceeding $3\sigma$ that correspond with identified outbursts. The data for the objects listed in Table~\ref{tab:novoutb} do not have three points exceeding $3\sigma$ anywhere in the light curve, with three exceptions: EPIC 211891315 shows evidence of a single possible outburst, which we describe in more detail in Section~\ref{sec:oDAV5}; EPIC 228682333 shows significant flux deviations in the first few points of the light curve which is likely the result of a poor reduction; EPIC 212100803---the faintest star in our sample---registers a brief sequence of three anomalously high points 39.11 days into observations immediately preceding a \Kep\ GO quality flag for simultaneous thruster firing, coarse point mode of the spacecraft, and a reaction wheel desaturation event. For these reasons, we do not accept the flagged points in the latter two objects as outbursts to the limits given in Table~\ref{tab:novoutb}.

\begin{figure*}
  \centering
  \includegraphics[width=0.99\columnwidth]{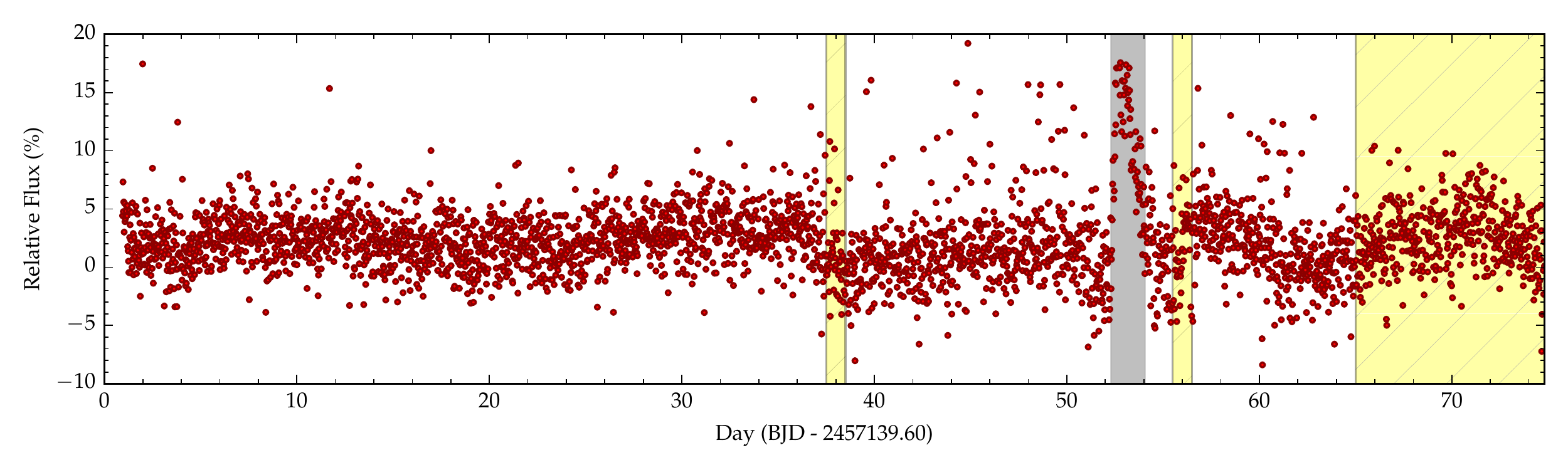}
  \caption{The long-cadence light curve of EPIC 211891315 ($K_p=19.4$\,mag) from \Ktwo. We highlight with solid gray the single feature starting near Day 52.3 that looks compellingly like an astrophysical brightening event. The three yellow hatch regions indicate identified instrumental systematics corresponding to (in chronological order): an argabrightening event, a likely CME, and a local background flux enhancement (see text).}
  \label{fig:oDAV5lc}
\end{figure*}

\begin{figure}
  \centering
  \includegraphics[width=0.99\columnwidth]{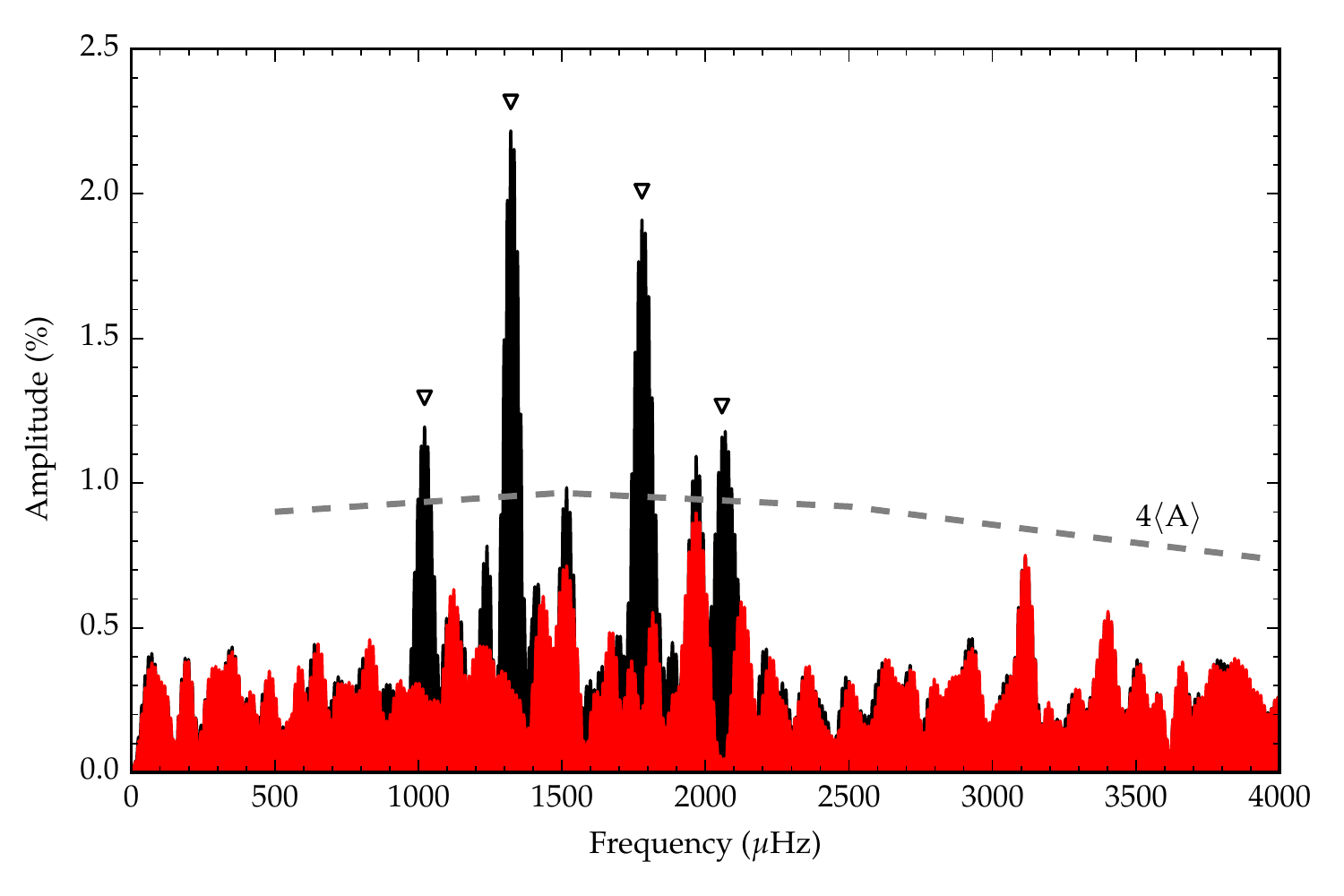}
  \caption{We confirmed pulsations in ground-based observations of the candidate outbursting white dwarf EPIC\,211891315. We show the original FT in black, and the prewhitened FT in red, after subtracting the four significant pulsation frequencies that are marked with triangles at 1322.0, 1780, 2057.7, and 1021.7 \muhz . The dashed grey line marks the running 4$\langle {\rm A}\rangle$ significance threshold.}
  \label{fig:mcdft}
\end{figure}

In addition to these targets observed at long cadence, we have also inspected the \Kep\ light curves of 11 pulsating white dwarfs that had previously published spectroscopic parameters and were proposed for short-cadence observations by our team to study their oscillations. We will present asteroseismology on these targets in forthcoming work, but we can rule out outbursts to various limits in each of these DAVs by considering their long-cadence observations in a manner identical to the other objects. Their atmospheric parameters were determined in the same way as our other targets, and constraints on the presence of outbursts in these light curves are included at the bottom of Table~\ref{tab:novoutb}.

Figure~\ref{fig:paramspace} shows that outbursts are narrowly confined to the lowest-temperature region of the empirical DAV instability strip between roughly $11{,}300$\,K and $10{,}600$\,K, below which pulsations are no longer observed. None of the other spectroscopically confirmed DAs observed by {\em K2} with temperatures outside this plot range show obvious outbursts, either.

Notably, there are two pulsating white dwarfs that have effective temperatures inside the region where we have detected the four other outbursting DAVs. 

KIC\,4357037, with $10{,}950 \pm 130$\,K and \logg{} = $8.11 \pm 0.04$ determined from WHT spectroscopy \citep{Greiss2016}, was observed continuously for 36.3\,d in the original \Kep\ mission, but did not show outbursts to a limit of at least 0.8\%. However, the pulsation spectrum of this white dwarf does not resemble a cool white dwarf, let alone the outbursting DAVs, with weighted mean pulsation period (WMP $= \sum{A_i P_i}/\sum{A_i}$, where $A_i$ are the measured amplitudes corresponding to $P_i$, the measured periods) of roughly 342.4\,s. WMPs systematically increase as white dwarfs cool and develop deeper convection zones, with values near 342.4\,s typically observed in white dwarfs with \teff\ $> 11,650$\,K (see, e.g., \citealt{Mukadam2006}, where the ``BG04'' sample from \citealt{Bergeron2004} and \citealt{Gianninas2005} is most comparable with the parameters derived in this work).
The interloper KIC\,4357037 may thus be hotter in actuality than its spectroscopic temperature suggests.

EPIC 60017836 \citep[also known as GD 1212; \teff\ $ = 10{,}970 \pm 170$\,K; \logg\ $ = 8.03 \pm 0.05$;][]{Hermes2014}, was observed for 9.0 continuous days during engineering time in preparation for \Ktwo\ operations.  With a \Kep\ magnitude of 13.3, we can rule out outbursts to an amplitude limit of 0.2\% that recur on timescales $\lesssim 9.0$\,d. The pulsation spectrum of this star is qualitatively similar to those of the outbursting DAVs, with a cluster of modes between $800-1270\, \mu$Hz, and additional significant peaks at higher frequency.  EPIC 60017836 is scheduled to be re-observed by \Ktwo\ in Campaign 12, which will allow us to explore the possibility of outbursts from this target with recurrence timescales $\gtrsim 9$\,d.

\subsection{EPIC 211891315: A Possible Single Outburst}
\label{sec:oDAV5}

The \Ktwo\ photometry for many of these faint objects is affected by long-term systematics, which are often aperiodic variations with timescales of several days. These trends very often also show up in the background flux, which is the median value of the background pixels that lie outside of the aperture used to extract the target photometry. These long-term trends often arise from solar coronal mass ejections (CMEs), small spacecraft thermal variations, so-called argabrightenings (see \Kep\ data release notes\footnote{\url{http://keplerscience.arc.nasa.gov/k2-data-release-notes.html}}), and electronic artifacts, such as rolling bands caused by time-varying crosstalk (e.g., \citealt{Clarke2014}).

In our inspection of the 52 DA white dwarfs within 2000\,K of $10{,}900$\,K, we found one object with a light curve that shows a significant brightening event that does not correlate with changes in other light curves on the same CCD module. The long-cadence \Ktwo\ Campaign 5 light curve of that white dwarf, EPIC\,211891315 ($K_p=19.4$\,mag, SDSSJ\,090231.76+183554.9), is shown in Figure~\ref{fig:oDAV5lc}.

There are two noted data anomalies with \Ktwo\ data taken in Campaign\,5, neither of which correlate with our observed brightness increase. The first, an unexplained argabrightening event, occurred roughly 38\,d into the campaign. The second, an increase in the median dark current likely caused by a CME, lasted for roughly one day starting 55.5\,d into the campaign.

We note that the brightening event we tentatively categorize as an outburst between Days $52.3-54.1$ does not correlate with any trends in the background flux, whereas the brightening at the end of the light curve between Days $65-75$ is correlated with an increase in background flux, strongly suggesting it is instrumental. We also rule out an asteroid or other contaminant moving through the photometric aperture during Days $52.3-54.1$ by inspecting the raw pixel data with the {\sc K2flix} visualization tool \citep{K2flix}. We tentatively identify EPIC\,211891315 as a candidate outbursting DAV. Our fit to the SDSS spectrum indicates this is a $11{,}310 \pm 410$\,K and \logg{} = $8.03 \pm 0.16$ white dwarf, included as an orange triangle in Figure~\ref{fig:paramspace}.

If this is an outburst of the same nature as in other DAVs observed by \Kep , it is by far the most energetic ever observed with an equivalent duration of 4.58\,hr, corresponding to an approximate total energy output of $1.5\times 10^{35}$\,erg.

Because the exposure time of long-cadence \Ktwo\ light curves is much longer than typical DAV pulsation periods, we followed this target up with high-speed photometry using the ProEM Camera on the 2.1m Otto Struve Telescope at McDonald Observatory. We observed EPIC\,211891315 for 4\,hr on the night of 2015~December~17 with 14\,s exposures through a broadband $BG40$ filter, which cuts off light redward of 6000 \AA\ to reduce sky noise. We obtained another 5\,hr of 10\,s exposures on 2015~December~18 through the $BG40$ filter.

The frames were dark subtracted and flat-fielded with standard {\sc IRAF} tasks. Aperture photometry was measured with the {\sc ccd\_hsp} package that utilizes {\sc IRAF} tasks from the {\sc phot} package \citep{Kanaan2002}. We used the {\sc WQED} software tools \citep{Thompson2013} to divide the flux measured for the target by the normalized flux from comparison stars in the field to correct for transparency variations, then divided each night's light curve by a second-order polynomial to account for differential airmass effects on stars with different colors. We repeated this process using a range of circular aperture sizes and adopt the apertures that yield the highest signal-to-noise. The FT of these two nights of data is displayed in Figure~\ref{fig:mcdft}.

With a total of 9\,hr of ground-based, time-series photometry over two nights, we were able to cleanly prewhiten (fit and subtract) significant sinusoidal signals from these data. For this reason, we adopt a different significance criterion for this data set. We measure the mean local amplitude along the FT, $\langle {\rm A}\rangle$, and use the standard 4$\langle {\rm A}\rangle$ significance threshold for single-site ground-based photometry \citep{Breger1993}.

We identify significant signals through an iterative process of calculating the FT and significance threshold, prewhitening all significant peaks, then recalculating the FT and threshold for the prewhitened light curve. We repeat this until no additional significant peaks are identified. Figure~\ref{fig:mcdft} displays the FTs of the original (black) and prewhitened (red) data with the final significance threshold and four significant frequencies marked. We establish EPIC\,211891315 as a new DAV. The four significant pulsation modes that we detect have frequencies $1322.0 \pm 0.4$ (2.22\% amplitude), $1779.0 \pm 0.4$ (1.99\%), $2057.7 \pm 0.7$ (1.19\%), and $1021.7 \pm 0.7\ \mu$Hz (1.12\%).

The confirmation of pulsations in this white dwarf provides marginal supporting evidence that the single brightening event during \Ktwo\ observations was a bona fide outburst. However, given the prevalence of systematic artifacts in the \Ktwo\ light curves of such faint targets, we are not comfortable confirming that EPIC\,211891315 is a new outbursting white dwarf, especially with only one event detected. Additionally, the observed pulsations have overall much higher frequencies than what we have measured in the four confirmed outbursting DAVs, with a weighted mean period of 685.9\,s. For now, we classify EPIC\,211891315 as only a candidate outbursting white dwarf.

\section{Discussion and conclusions}
\label{sec:conc}

The four confirmed members of the outbursting class of DAV have three distinct commonalities: (1) repeated outbursts, recurring on irregular intervals of order days and lasting for several hours;
(2) effective temperatures that put them near the cool, red edge of the DAV instability strip; and (3) rich pulsation spectra dominated by low-frequency ($800-1400$\,s period) pulsations that are unstable in amplitude/frequency with at least one stable mode at significantly higher frequency ($350-515$\,s, and maybe as short as 290\,s), which in the first two cases appeared to be an $\ell=1$ from rotational splittings \citep{Bell2015,Hermes2015}. We summarize their main characteristics in Table~\ref{tab:summary}.

\begin{deluxetable*}{c c c c c c c c c}[t]
\tablecolumns{9}
\tablecaption{Properties of Outbursting DAVs \label{tab:summary}}
\tablehead{
\colhead{Name} & \colhead{$K_p$} & \colhead{\teff}	& \colhead{\logg} &  \colhead{$\tau_{recur}$}  &  \colhead{Med. Duration}  &  \colhead{Max. Flux} & \colhead{Max. Energy} &  \colhead{Reference} \\  \colhead{} & \colhead{(mag)} & \colhead{(K)}	& \colhead{(cgs)} &  \colhead{(d)}  &  \colhead{(hr)}  &  \colhead{(\%)} &  \colhead{(erg)} &  \colhead{}   }
\startdata
KIC\,4552982    & 17.9 & $10{,}860(120)$ & $8.16(0.06)$ & 2.7 & 9.6 & 17  & $2.1\times 10^{33}$ & \citet{Bell2015} \\
PG\,1149+057    & 15.0 & $11{,}060(170)$ & $8.06(0.05)$ & 8.0 &  15 & 45 & $1.2\times 10^{34}$ & \citet{Hermes2015} \\
EPIC\,211629697 & 18.4 & $10{,}570(120)$ & $7.92(0.07)$ & 5.0 & 16.3 & 15 & $1.8\times 10^{34}$ & This work \\
EPIC\,229227292 & 16.7 & $11{,}190(170)$ & $8.02(0.05)$ & 2.4 & 10.2 & 9 & $3.1\times 10^{33}$ & This work 
\enddata
\end{deluxetable*}

The discovery of repeated outbursts in four of the first 16 DAVs observed by the {\em Kepler} spacecraft indicates that this is not an incredibly rare phenomenon. However, it does beg the question of how outbursts have been missed during the first 45 years of studies of pulsating white dwarfs.

In this context, the minimum outburst duration observed offers a clue: So far, every outburst lasts for more than several hours. Nearly all previous ground-based, time-series photometry of pulsating white dwarfs involves differential photometry: dividing the target by a (usually redder) comparison star to compensate for changing atmospheric conditions. Due to color-dependent extinction effects, nearly all groups have adopted a methodology of dividing out at least a second-order polynomial to normalize the light curves \citep[e.g.][]{Nather1990}. It is possible that outbursts were observed during previous ground-based studies of pulsating white dwarfs but were unintentionally de-trended from the data. Notably, the DBV (pulsating helium-atmosphere white dwarf) GD 358 underwent a large-scale brightening event in 1996, which may have been the first documented case of an outburst in a pulsating white dwarf \citep{Nitta1999,Montgomery2010}.

The physical mechanism that causes outbursts remains an exciting open question. \citet{Hermes2015} suggested that, following the theoretical framework laid out by \citet{Wu2001}, the outbursts could be the result of nonlinear three-mode resonant coupling. In this model, energy is transferred from an observed, overstable parent mode to daughter modes via parametric resonance, one or both of which may be damped by turbulence in the convection zone and deposit their newfound energy there.

All four of the outbursting white dwarfs have some of the longest pulsation periods observed in DAVs, excluding the extremely low-mass white dwarfs \citep{Hermes2013}. \citet{Wu2001} predicted that mode coupling would be most prevalent in the coolest white dwarfs with the longest-period pulsations, simply because there are more possible modes with which to couple.

By inspecting the light curves of the more than 300 spectroscopically confirmed DA white dwarfs observed already by {\em K2}, we have shown that outbursts only occur in a narrow temperature range, between roughly $11{,}300$\,K and $10{,}600$\,K. This temperature range falls just hot of the empirical red edge of the DAV instability strip, below which pulsations are no longer observed.

The red edge of the DAV instability strip has been notoriously difficult to predict from nonadiabatic pulsation codes, which suggest that white dwarfs should have observable pulsations down to at least 6000\,K \citep[e.g.][]{VanGrootel2012}. There have been two proposed mechanisms to bring the theoretical red edge in line with observations.

\citet{Hansen1985} suggested that there is a critically maximal mode period, beyond which $g$-modes are no longer reflected off the outer mode cavity and thus evanesce. \citet{VanGrootel2013} showed that applying this critical mode period for $\ell=1$ modes to the thermal timescale at the base of the convection zone can successfully reproduce the empirical red edge of the DAV instability strip across a wide range of white dwarf masses.

Additionally, a series of papers by Wu \& Goldreich proposed amplitude saturation mechanisms in the coolest DAVs from turbulent viscosity of the convection zone as well as resonant three-mode interactions as ways to cause a hotter red edge than nonadiabatic predictions \citep{Goldreich1999b,Wu2001}. If outbursts are indeed caused by nonlinear mode coupling, this suggests amplitude saturation as an important contributor to the cessation of observability of pulsations in the coolest DAVs.

The measured properties of outbursts provide observational leverage for efforts to understand pulsational mode selection and driving, especially in the context of the few short-period modes that are selected in all four of the outbursting DAVs. Fortunately, DAV pulsations are extremely sensitive to structural changes in white dwarfs, and our understanding of outbursts will benefit from further asteroseismic analysis of these objects that will be the subject of future work.

\Ktwo\ continues to obtain extensive space-based photometry on new fields roughly every three months, and we look forward to inspecting future data releases for additional instances of this exciting physical phenomenon.

\acknowledgements

We thank the referee, S.~O.~Kepler, for comments that helped us to improve this manuscript. We thank Geert Barentsen and the \Ktwo\ Guest Observer office for their quick investigation of charge bleed from m Vir, and for D.~J.~Armstrong for use of his extraction and detrending pipeline. We also thank the teams led by M.~Kilic, M.~R.~Burleigh, Seth~Redfield, Avi~Shporer, R.~Alonso, and Steven~D.~Kawaler for securing \Ktwo\ observations on a wide variety of white dwarfs, proposing many of the white dwarfs discussed here.
K.J.B., M.H.M., D.E.W., and K.I.W. acknowledge support from NSF grant AST-1312983, the \Kep{} Cycle 4 GO proposal 11-KEPLER11-0050, and NASA grant NNX13AC23G. Support for this work was provided by NASA through Hubble Fellowship grant \#HST-HF2-51357.001-A, awarded by the Space Telescope Science Institute, which is operated by the Association of Universities for Research in Astronomy, Incorporated, under NASA contract NAS5-26555. The research leading to these results has received funding from the European Research Council under the European Union's Seventh Framework Programme (FP/2007-2013) / ERC Grant Agreement n. 320964 (WDTracer).  A.G. gratefully acknowledges the support of the NSF under grant AST-1312678, and NASA under grant NNX14AF65G.
The short-cadence \Ktwo\ data were obtained thanks to Guest Observer programs in Cycle 1 (GO5043) and Cycle 2 (GO6083).  This paper includes data collected by the \Kep\ mission. Funding for the \Kep\ mission is provided by the NASA Science Mission directorate. Some of the data presented in this paper were obtained from the Mikulski Archive for Space Telescopes (MAST). STScI is operated by the Association of Universities for Research in Astronomy, Inc., under NASA contract NAS5-26555. Support for MAST for non-HST data is provided by the NASA Office of Space Science via grant NNX09AF08G and by other grants and contracts. This paper includes data taken at The McDonald Observatory of The University of Texas at Austin, as well as observations obtained at the Southern Astrophysical Research (SOAR) telescope, which is a joint project of the Minist\'{e}rio da Ci\^{e}ncia, Tecnologia, e Inova\c{c}\~{a}o (MCTI) da Rep\'{u}blica Federativa do Brasil, the U.S. National Optical Astronomy Observatory (NOAO), the University of North Carolina at Chapel Hill (UNC), and Michigan State University (MSU). The authors acknowledge the Texas Advanced Computing Center (TACC) at The University of Texas at Austin for providing data archiving resources that have contributed to the research results reported within this paper.

\begin{deluxetable*}{l r r r r r r r c}[h]
\tablecolumns{9}
\tablecaption{White Dwarfs Not Observed to Outburst with \Kep\ Observations \label{tab:novoutb}}
\tablehead{
\colhead{EPIC ID} & \colhead{$K_p$ (mag)} & \colhead{\teff\ (K)}	& \colhead{\logg\ (cgs)} &  \colhead{Pipe.\tablenotemark{a}} & \colhead{Field}  & \colhead{Obs. Dur. (d)} & \colhead{Limit (\%)} & Ref.\tablenotemark{b} }
\startdata
212154350 & 19.8 & 12900(790) & 7.91(0.21) & VJ & 5 & 73.9 & 8.1 & 1 \\
212100803 & 20.0 & 12600(650) & 7.80(0.21) & GO & 5 & 74.8 & 16.7 & 1 \\
206284230 & 19.0 & 12570(430) & 8.03(0.11) & GO & 3 & 69.1 & 2.5 & 1 \\
210484300\tablenotemark{c} & 19.0 & 12490(450) & 8.52(0.09) & VJ & 4 & 68.6 & 2.9 & 1 \\
211975984 & 19.3 & 12480(440) & 8.08(0.11) & VJ & 5 & 73.9 & 5.6 & 1 \\
211888384 & 18.3 & 12330(210) & 8.33(0.05) & GO & 5 & 74.8 & 1.3 & 1 \\
228682421 & 19.7 & 12070(450) & 8.17(0.13) & VJ & 5 & 73.9 & 6.1 & 1 \\
211564222 & 19.8 & 12060(510) & 8.07(0.16) & VJ & 5 & 73.9 & 11.6 & 1 \\
211934410 & 19.0 & 11940(280) & 7.72(0.11) & VJ & 5 & 73.9 & 3.4 & 1 \\
201754145 & 19.4 & 11930(460) & 8.05(0.15) & VJ & 1 & 80.1 & 3.4 & 1 \\
201331010 & 19.3 & 11670(290) & 8.16(0.09) & VJ & 1 & 80.1 & 2.8 & 1 \\
228682407 & 19.9 & 11620(680) & 8.24(0.21) & VJ & 5 & 73.9 & 13.6 & 1 \\
228682371 & 19.9 & 11420(480) & 8.22(0.17) & VJ & 5 & 73.9 & 8.2 & 1 \\
228682428 & 19.8 & 11380(610) & 8.52(0.22) & GO & 5 & 74.8 & 11.1 & 1 \\
211891315 & 19.4 & 11310(410) & 8.03(0.16) & VJ & 5 & 73.9 & n/a\tablenotemark{d} & 1 \\
228682357 & 19.9 & 11240(390) & 8.09(0.15) & VJ & 5 & 73.9 & 6.4 & 1 \\
211330756 & 19.9 & 11090(600) & 8.05(0.27) & GO & 5 & 74.8 & 14.3 & 1 \\
201259883 & 19.7 & 11010(450) & 8.17(0.21) & VJ & 1 & 80.1 & 8.0 & 1 \\
228682400 & 19.9 & 10970(410) & 8.13(0.20) & VJ & 5 & 73.9 & 22.7 & 1 \\
212169533 & 19.9 & 10890(460) & 7.90(0.23) & GO & 5 & 74.8 & 15.6 & 1 \\
212091315 & 19.4 & 10630(210) & 8.39(0.14) & VJ & 5 & 73.8 & 3.7 & 1 \\
211886776 & 19.0 & 10570(150) & 8.09(0.09) & VJ & 5 & 73.9 & 2.1 & 1 \\
203705962 & 15.1 & 10380(120) & 7.95(0.09) & VJ & 2 & 77.5 & 0.2 & 2 \\
228682361 & 19.8 & 10340(220) & 7.95(0.17) & VJ & 5 & 73.9 & 4.2 & 1 \\
212071753 & 18.9 & 10300(150) & 8.11(0.12) & VJ & 5 & 73.9 & 3.8 & 1 \\
206302487 & 18.7 & 10190(110) & 8.13(0.10) & VJ & 3 & 66.8 & 1.3 & 1 \\
211932844 & 17.9 & 10060(80) & 8.09(0.07) & VJ & 5 & 73.9 & 1.4 & 1 \\
211519519 & 18.9 & 9990(130) & 7.99(0.13) & VJ & 5 & 73.9 & 2.1 & 1 \\
228682409 & 20.0 & 9960(160) & 8.22(0.15) & GO & 5 & 74.8 & 13.5 & 1 \\
201513373 & 18.2 & 9800(80) & 8.13(0.08) & VJ & 1 & 80.0 & 0.9 & 1 \\
201789520 & 18.4 & 9710(80) & 7.88(0.09) & VJ & 1 & 80.0 & 1.2 & 1 \\
201498548 & 18.3 & 9680(60) & 8.07(0.07) & VJ & 1 & 80.0 & 1.5 & 1 \\
201663682 & 19.0 & 9530(110) & 8.07(0.12) & VJ & 1 & 80.1 & 4.5 & 1 \\
201810512 & 18.4 & 9510(90) & 8.20(0.09) & VJ & 1 & 80.0 & 1.7 & 1 \\
228682315 & 19.5 & 9470(150) & 7.74(0.19) & GO & 5 & 74.8 & 11.0 & 1 \\
201838978 & 18.7 & 9460(80) & 7.79(0.10) & VJ & 1 & 80.1 & 2.5 & 1 \\
211932489 & 19.8 & 9450(180) & 8.03(0.20) & VJ & 5 & 73.9 & 5.6 & 1 \\
201834393 & 18.7 & 9440(90) & 7.85(0.11) & VJ & 1 & 80.1 & 2.7 & 1 \\
201887383 & 18.8 & 9400(100) & 8.14(0.11) & VJ & 1 & 80.0 & 6.5 & 1 \\
201521421 & 19.2 & 9340(160) & 8.09(0.17) & VJ & 1 & 80.1 & 2.0 & 1 \\
201879492 & 18.1 & 9320(70) & 8.05(0.08) & VJ & 1 & 80.0 & 1.8 & 1 \\
201816218 & 18.4 & 9260(80) & 7.98(0.09) & VJ & 1 & 80.0 & 1.1 & 1 \\
211768391 & 18.5 & 9250(80) & 7.60(0.12) & VJ & 5 & 73.9 & 2.1 & 1 \\
211692110 & 18.8 & 9180(90) & 7.59(0.13) & VJ & 5 & 73.9 & 2.2 & 1 \\
201224667 & 18.6 & 9180(110) & 8.02(0.13) & VJ & 1 & 80.0 & 3.2 & 1 \\
201723220 & 17.7 & 9110(50) & 8.05(0.05) & VJ & 1 & 80.0 & 0.8 & 1 \\
211788137 & 18.6 & 9100(100) & 7.98(0.12) & VJ & 5 & 73.9 & 3.0 & 1 \\
228682387 & 18.8 & 9100(90) & 7.90(0.11) & VJ & 5 & 72.0 & 1.6 & 1 \\
201700041 & 19.2 & 9070(160) & 7.99(0.19) & VJ & 1 & 80.1 & 3.3 & 1 \\
212564858 & 15.7 & 9050(110) & 7.83(0.03) & GO & 6 & 78.9 & 0.3 & 3 \\
228682333 & 17.8 & 9000(60) & 7.76(0.09) & VJ & 5 & 73.9 & 2.6 & 1 \\
228682427 & 18.6 & 8980(90) & 8.17(0.10) & VJ & 5 & 73.9 & 4.9 & 1 \\
\multicolumn{9}{c}{Pulsating White Dwarfs Not Observed to Outburst with Short-Cadence \Kep{} Observations} \\
201730811 & 15.7 & 12490(260) & 8.01(0.06) & VJ & 1 & 80.1 & 1.8 & 4 \\
212395381 & 15.7 & 12020(190) & 8.18(0.05) & GO & 6 & 73.9 & 2.5 & 5 \\
10132702 & 19.1 & 11940(380) & 8.12(0.04) & GO & K1 & 30.8 & 2.5 & 6 \\
211916160 & 19.0 & 11900(230) & 8.23(0.07) & VJ & 5 & 73.9 & 2.1 & 1 \\
7594781 & 18.2 & 11730(140) & 8.11(0.04) & GO & K1 & 31.8 & 0.6 & 6 \\
211926430 & 17.7 & 11690(120) & 8.09(0.04) & VJ & 5 & 73.9 & 3.1 & 1 \\
11911480 & 18.1 & 11580(140) & 7.96(0.04) & GO & K1 & 82.6 & 1.9 & 6 \\
201719578 & 18.1 & 10990(125) & 7.91(0.06) & VJ & 1 & 80.1 & 2.1 & 1 \\
211596649 & 19.0 & 11230(260) & 7.94(0.11) & VJ & 5 & 73.8 & 2.4 & 1 \\
60017836 & 13.3 & 10970(170) & 8.03(0.05) & GO & Eng & 9.0 & 0.2 & 5 \\
4357037 & 18.3 & 10950(130) & 8.11(0.04) & GO & K1 & 36.3 & 0.8 & 4 \\
\tablenotetext{a}{\Ktwo\ reduction pipeline, where GO is the \Kep\ Guest Observer light curve, and VJ is the \citet{Vanderburg2014} optimally extracted light curve.}
\tablenotetext{b}{Spectroscopic sources: (1)~\citet{Tremblay2011}; (2)~\citet{Kawka2006}; (3)~\citet{Koester2009}; (4)~\citet{Hermes2015a}; (5)~\citet{Gianninas2011}; (6)~\citet{Greiss2016}}
\tablenotetext{c}{The atmosphere model that best fits the Balmer line profiles of the spectrum of EPIC\,210484300 disagrees with its photometric colors from SDSS.}
\tablenotetext{d}{EPIC 211891315 shows evidence of a single outburst and was observed to pulsate in follow-up, ground-based observations as discussed in Section \ref{sec:oDAV5}.}

\enddata
\end{deluxetable*}


\begin{thebibliography}{}

\bibitem[Althaus et al.(2010)]{Althaus2010} Althaus, L.~G., C{\'o}rsico, A.~H., Isern, J., \& Garc{\'{\i}}a-Berro, E.\ 2010, \aapr, 18, 471 
\bibitem[Armstrong et al.(2015)]{Armstrong2015} Armstrong, D.~J., Kirk, J., Lam, K.~W.~F., et al.\ 2015, \aap, 579, A19 
\bibitem[Baran et al.(2015)]{Baran2015} Baran, A.~S., Koen, C., \& Pokrzywka, B.\ 2015, \mnras, 448, L16 
\bibitem[Barentsen(2015)]{K2flix} Barentsen, G.\ 2015, Astrophysics Source Code Library, ascl:1503.001 
\bibitem[Bell et al.(2015)]{Bell2015} Bell, K.~J., Hermes, J.~J., Bischoff-Kim, A., et al.\ 2015, \apj, 809, 14 
\bibitem[Bergeron et al.(2004)]{Bergeron2004} Bergeron, P., Fontaine, G., Bill{\`e}res, M., Boudreault, S., \& Green, E.~M.\ 2004, \apj, 600, 404 
\bibitem[Breger et al.(1993)]{Breger1993} Breger, M., Stich, J., Garrido, R., et al.\ 1993, \aap, 271, 482 
\bibitem[Brickhill(1991)]{Brickhill1991} Brickhill, A.~J.\ 1991, \mnras, 251, 673 
\bibitem[Clarke et al.(2014)]{Clarke2014} Clarke, B., Kolodziejczak, J.~J, \& Caldwell, D.~A.\ 2014, American Astronomical Society Meeting Abstracts \#224, 224, 120.07 
\bibitem[Clemens et al.(2004)]{Clemens2004} Clemens, J.~C., Crain, J.~A., \& Anderson, R.\ 2004, \procspie, 5492, 331 
\bibitem[Hansen et al.(1985)]{Hansen1985} Hansen, C.~J., Winget, D.~E., \& Kawaler, S.~D.\ 1985, \apj, 297, 544 
\bibitem[Hermes et al.(2011)]{Hermes2011} Hermes, J.~J., Mullally, F., {\O}stensen, R.~H., et al.\ 2011, \apjl, 741, L16 
\bibitem[Hermes et al.(2013)]{Hermes2013} Hermes, J.~J., Montgomery, M.~H., Gianninas, A., et al.\ 2013, \mnras, 436, 3573 
\bibitem[Hermes et al.(2014)]{Hermes2014} Hermes, J.~J., Charpinet, S., Barclay, T., et al.\ 2014, \apj, 789, 85 
\bibitem[Hermes et al.(2015a)]{Hermes2015a} Hermes, J.~J., G{\"a}nsicke, B.~T., Bischoff-Kim, A., et al.\ 2015a, \mnras, 451, 1701 
\bibitem[Hermes et al.(2015b)]{Hermes2015} Hermes, J.~J., Montgomery, M.~H., Bell, K.~J., et al.\ 2015b, \apjl, 810, L5 
\bibitem[Holberg \& Bergeron(2006)]{Holberg2006} Holberg, J.~B., \& Bergeron, P.\ 2006, \aj, 132, 1221 
\bibitem[Howell et al.(2014)]{Howell2014} Howell, S.~B., Sobeck, C., Haas, M., et al.\ 2014, \pasp, 126, 398 
\bibitem[Fontaine et al.(2001)]{Fontaine2001} Fontaine, G., Brassard, P., \& Bergeron, P.\ 2001, \pasp, 113, 409 
\bibitem[Fontaine \& Brassard(2008)]{Fontaine2008} Fontaine, G., \& Brassard, P.\ 2008, \pasp, 120, 1043 
\bibitem[Gianninas et al.(2005)]{Gianninas2005} Gianninas, A., Bergeron, P., \& Fontaine, G.\ 2005, \apj, 631, 1100 
\bibitem[Gianninas et al.(2011)]{Gianninas2011} Gianninas, A., Bergeron, P., \& Ruiz, M.~T.\ 2011, \apj, 743, 138
\bibitem[Gilliland et al.(2010)]{Gilliland2010} Gilliland, R.~L., Jenkins, J.~M., Borucki, W.~J., et al.\ 2010, \apjl, 713, L160 
\bibitem[Goldreich \& Wu(1999a)]{Goldreich1999} Goldreich, P., \& Wu, Y.\ 1999a, \apj, 511, 904 
\bibitem[Goldreich \& Wu(1999b)]{Goldreich1999b} Goldreich, P., \& Wu, Y.\ 1999b, \apj, 523, 805 
\bibitem[Greiss et al.(2016)]{Greiss2016} Greiss, S., Hermes, J.~J., G{\"a}nsicke, B.~T., et al.\ 2016, \mnras, 457, 2855 
\bibitem[Horne(1986)]{Horne1986} Horne, K.\ 1986, \pasp, 98, 609 
\bibitem[Kanaan et al.(2002)]{Kanaan2002} Kanaan, A., Kepler, S.~O., \& Winget, D.~E.\ 2002, \aap, 389, 896 
\bibitem[Kawka \& Vennes(2006)]{Kawka2006} Kawka, A., \& Vennes, S.\ 2006, \apj, 643, 402 
\bibitem[Koester et al.(2009)]{Koester2009} Koester, D., Voss, B., Napiwotzki, R., et al.\ 2009, \aap, 505, 441 
\bibitem[Kowalski \& Saumon(2006)]{Kowalski2006} Kowalski, P.~M., \& Saumon, D.\ 2006, \apjl, 651, L137 
\bibitem[Kleinman et al.(2013)]{Kleinman2013} Kleinman, S.~J., Kepler, S.~O., Koester, D., et al.\ 2013, \apjs, 204, 5 
\bibitem[Lenz \& Breger(2004)]{Lenz2004} Lenz, P., \& Breger, M.\ 2004, The A-Star Puzzle, 224, 786 
\bibitem[Marsh(1989)]{Marsh1989} Marsh, T.~R.\ 1989, \pasp, 101, 1032 
\bibitem[Montgomery \& Odonoghue(1999)]{Montgomery1999} Montgomery, M.~H., \& Odonoghue, D.\ 1999, Delta Scuti Star Newsletter, 13, 28 
\bibitem[Montgomery et al.(2010)]{Montgomery2010} Montgomery, M.~H., Provencal, J.~L., Kanaan, A., et al.\ 2010, \apj, 716, 84 
\bibitem[Mukadam et al.(2006)]{Mukadam2006} Mukadam, A.~S., Montgomery, M.~H., Winget, D.~E., Kepler, S.~O., \& Clemens, J.~C.\ 2006, \apj, 640, 956 
\bibitem[Mukadam et al.(2007)]{Mukadam2007} Mukadam, A.~S., Montgomery, M.~H., Kim, A., et al.\ 2007, 15th European Workshop on White 
Dwarfs, 372, 587 
\bibitem[Nather et al.(1990)]{Nather1990} Nather, R.~E., Winget, D.~E., Clemens, J.~C., Hansen, C.~J., \& Hine, B.~P.\ 1990, \apj, 361, 309 
\bibitem[Nitta et al.(1999)]{Nitta1999} Nitta, A., Winget, D.~E., Kepler, S.~O., et al.\ 1999, 11th European Workshop on White Dwarfs, 169, 104 
\bibitem[Robinson et al.(1982)]{Robinson1982} Robinson, E.~L., Kepler, S.~O., \& Nather, R.~E.\ 1982, \apj, 259, 219 
\bibitem[Shanks et al.(2015)]{Shanks2015} Shanks, T., Metcalfe, N., Chehade, B., et al.\ 2015, \mnras, 451, 4238 
\bibitem[Thompson \& Mullally(2013)]{Thompson2013} Thompson, S., \& Mullally, F.\ 2013, Astrophysics Source Code Library, ascl:1304.004 
\bibitem[Tremblay \& Bergeron(2008)]{Tremblay2008} Tremblay, P.-E., \& Bergeron, P.\ 2008, \apj, 672, 1144 
\bibitem[Tremblay \& Bergeron(2009)]{Tremblay2009} Tremblay, P.-E., \& Bergeron, P.\ 2009, \apj, 696, 1755 
\bibitem[Tremblay et al.(2011)]{Tremblay2011} Tremblay, P.-E., Bergeron, P., \& Gianninas, A.\ 2011, \apj, 730, 128 
\bibitem[Tremblay et al.(2013)]{Tremblay2013} Tremblay, P.-E., Ludwig, H.-G., Steffen, M., \& Freytag, B.\ 2013, \aap, 559, A104 
\bibitem[Tremblay et al.(2015)]{Tremblay2015} Tremblay, P.-E., Gianninas, A., Kilic, M., et al.\ 2015, \apj, 809, 148 
\bibitem[Twicken et al.(2010)]{Twicken2010} Twicken, J.~D., Chandrasekaran, H., Jenkins, J.~M., et al.\ 2010, \procspie, 7740, 77401U 
\bibitem[Van Grootel et al.(2012)]{VanGrootel2012} Van Grootel, V., Dupret, M.-A., Fontaine, G., et al.\ 2012, \aap, 539, A87
\bibitem[Van Grootel et al.(2013)]{VanGrootel2013} Van Grootel, V., Fontaine, G., Brassard, P., \& Dupret, M.-A.\ 2013, \apj, 762, 57 
\bibitem[Vanderburg \& Johnson(2014)]{Vanderburg2014} Vanderburg, A., \& Johnson, J.~A.\ 2014, \pasp, 126, 948 
\bibitem[Vanderburg et al.(2015)]{Vanderburg2015} Vanderburg, A., Johnson, J.~A., Rappaport, S., et al.\ 2015, \nat, 526, 546  
\bibitem[Winget \& Kepler(2008)]{Winget2008} Winget, D.~E., \& Kepler, S.~O.\ 2008, \araa, 46, 157 
\bibitem[Wu \& Goldreich(2001)]{Wu2001} Wu, Y., \& Goldreich, P.\ 2001, \apj, 546, 469 


\end{thebibliography}
\end{document}